\documentclass[final, 5p, times]{elsarticle}
\DeclareUnicodeCharacter{2113}{\ensuremath{\ell}}
\usepackage{subcaption}
\usepackage{ulem}
\usepackage{xcolor}
\usepackage{amsmath}
\usepackage{etoolbox}
\makeatletter
\def\@alph#1{\number#1}
\def\ps@pprintTitle{%
 \let\@oddhead\@empty
 \let\@evenhead\@empty
 \let\@oddfoot\@empty
 \let\@evenfoot\@empty
}
\makeatother
\begin{document}
\begin{frontmatter}

\title{MINER Reactor Based Search for Axion-Like Particles Using Sapphire (\(\text{Al}_2\text{O}_3\)) Detectors}
\author[inst1]{M. Mirzakhani}
\author[inst1]{W. Baker}
\author[inst2]{M. Chaudhuri}
\author[inst3]{J. B. Dent}
\author[inst2]{R. Dey}
\author[inst1]{B. Dutta}
\author[inst4]{V. Iyer}
\author[inst1]{A. Jastram}
\author[inst2]{V. K. S. Kashyap}
\author[inst5]{A. Kubik}
\author[inst6]{K. Lang}
\author[inst1]{R. Mahapatra}
\author[inst1]{S. Maludze}
\author[inst1]{N. Mirabolfathi}
\author[inst2]{B. Mohanty}
\author[inst2]{D. Mondal}
\author[inst7]{H. Neog}
\author[inst8]{J.~L.~Newstead}
\author[inst1]{M. Platt}
\author[inst1]{S. Sahoo}
\author[inst9]{J. Sander}
\author[inst1]{L. E. Strigari}
\author[inst3]{J. Walker}

\affiliation[inst1]{organization={Department of Physics and Astronomy, Texas A\&M University },
            addressline={578 University Dr}, 
            city={College Station},
            postcode={77840}, 
            state={TX},
            country={US}}
            
\affiliation[inst2]{organization={National Institute of Science Education and Research, An OCC of Homi Bhabha National Institute, Jatni  752050, India}}
\affiliation[inst3]{organization={Department of Physics, Sam Houston State University },
            addressline={1905 University Ave}, 
            city={Huntsville}, 
            state={TX},
            postcode={77340},
            country={US}}

\affiliation[inst4]{organization={Department of Physics, University of Toronto},
            addressline={27 King's College Cir}, 
            city={Toronto},
            postcode={ON M5S 1A7}, 
            country={Canada}}
\affiliation[inst5]{SNOLAB, Creighton Mine \#9, 1039 Regional Road 24, Sudbury, ON P3Y 1N2, Canada}
\affiliation[inst6]{organization={Department of Physics, The University of Texas at Austin},
            addressline={2515 Speedway},
            city={Austin}, 
            state={TX},
            postcode={78712},
            country={US}}
\affiliation[inst7]{organization={School of Physics \& Astronomy, University of Minnesota},
            city={Minneapolis},
            state={MN},
            postcode={55455}, 
            country={US}}
\affiliation[inst8]{organization={ARC Centre of Excellence for Dark Matter Particle Physics, School of Physics, The University of Melbourne},
            addressline={Parkville VIC 3010},
            country={Australia}}

\affiliation[inst9]{organization={Department of Physics, University of South Dakota},
            addressline={414 E Clark St},
            city={Vermillion},
            state={SD},
            postcode={57069}, 
            country={US}}

\begin{abstract}
The absence of definitive results for WIMP dark matter has sparked growing interest in alternative dark matter candidates (ALPs), such as axions and Axion-Like Particles, which also provide insight into the strong CP problem. The Mitchell Institute Neutrino Experiment at Reactor (MINER), conducted at the Nuclear Science Center of Texas A\&M University, investigated ALPs near a 1 MWth TRIGA nuclear reactor core, positioned approximately 4 meters away. This experiment employed cryogenic Sapphire detectors with a low detection threshold ($\approx$ 100 eV
), equipped with a Transition Edge Sensor capable of detecting athermal phonons. Due to the low-background environment, we were able to exclude ALPs with axion-photon coupling and axion-electron coupling as small as $g_{a\gamma\gamma} = 10^{-5}$ and  $g_{aee} = 10^{-7}$, respectively. Energy depositions below 3\,keV were not considered and remain blinded for our Coherent Elastic Neutrino Nucleus Scattering (CE$\nu$NS) analysis. This is the first result demonstrating the MINER experiment's potential to probe a low-mass ALPs, enabled by its low-threshold detector and proximity to a reactor.

\end{abstract}
\end{frontmatter}

\section{Introduction}

One of the most compelling solutions to the Strong CP problem involves introducing a new broken symmetry, known as the Peccei-Quinn (PQ) symmetry, which naturally and dynamically cancels the CP violation in the strong interaction~\cite{Peccei, Peccei1, mirzakhani}. A direct consequence of this symmetry is the emergence of a hypothetical pseudoscalar particle called the axion~\cite{Weinberg, Wilczek}. Axions and their generalizations, Axion-Like Particles (ALPs), have attracted significant experimental interest due to their potential role in particle physics and cosmology.

Experimental searches for axions and ALPs leverage their coupling to photons in a variety of setups. Helioscope experiments such as CAST~\cite{zioutas, cast} and the future IAXO~\cite{irastorza} seek axions emitted by the Sun, while haloscope experiments such as ADMX~\cite{Asztalos, Du}, Abracadabra~\cite{kahn, salemi}, and HAYSTAC~\cite{Brubaker, droster} aim to detect axions within a strong magnetic field. Light-shining through-wall experiments such as ALPS II~\cite{spector} attempt to produce and regenerate axions under controlled laboratory conditions. Furthermore, solar axions produced through nuclear transitions can be constrained by resonant absorption in laboratory nuclei~\cite{Moriyama, KRCMAR, KRCMAR1, Derbin, Creswick, Gangapshev}.

\begin{figure}[h!]
  \centering
    \includegraphics[width=\linewidth]{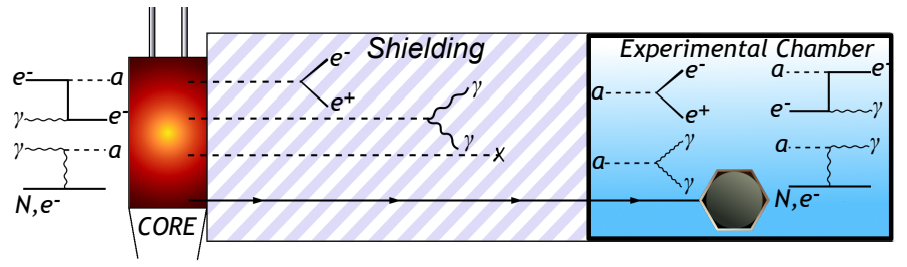}
   \caption{A schematic illustration of axion-like particles (ALPs) and their interactions at a reactor neutrino experiment. On the left, the production of ALPs is shown, while the right highlights their scattering and decay possibilities. ALPs may decay inside the reactor shielding, evading detection (dashed lines). Alternatively, ALPs that free stream through the shielding (solid lines) can be detected via the inverse Primakoff process, inverse Compton scattering channels, or decay channels.}
  \label{fig:fig1}
\end{figure}

Reactor-based experiments provide a unique opportunity for axion searches, since they offer a controlled environment for both axion production and detection. The MINER experiment at Texas A\&M University employs an array of low-threshold cryogenic sapphire detectors placed 4 meters from a 1 MWth TRIGA reactor core~\cite{minercollaboration}. This setup enables the search for ALPs produced via photon interactions within the reactor tank. ALPs can be detected through their scattering off nuclei and electrons in the detector or via their decay into photons or electron-positron pairs, allowing constraints on both axion-photon and axion-electron couplings. A schematic of the experimental setup is shown in Fig.~\ref{fig:fig1}. Similar reactor-based experiments such as CONUS~\cite{buck}, CONNIE~\cite{conniecollaboration}, and $\nu$-cleus~\cite{Strauss} employ different detector technologies to probe ALPs. Although astrophysical observations also provide constraints on the ALP parameter space, laboratory experiments offer a direct and controlled alternative, avoiding uncertainties associated with astrophysical modeling.

This paper is structured as follows. Section~\ref{Theory} presents the theoretical framework, including the ALP production mechanisms in the reactor, the detection methods, and the experimental sensitivity. Section~\ref{Experimental} describes the experimental setup, detailing the reactor, detector configuration, data acquisition process, and shielding techniques. In Section~\ref{Analysis}, we present the experimental results, followed by an exclusion limit on ALP searches in the relevant parameter space. Finally, in Section~\ref{Conclusion}, we discuss our conclusions and future improvements to enhance the sensitivity of the experiment and to extend its reach to unexplored regions of the ALP parameter space.
 
\section{ALP production and detection} \label{Theory}

As the primary goal of this experiment is to focus on the axion-photon and axion-electron coupling, we consider a generic model where the ALP interacts with photons or electrons, as described by the interaction terms in the Lagrangian:
\begin{equation} \label{eq:lagrangian}
\mathcal{L}_{\text{int}} \supset -\frac{1}{4}g_{a\gamma\gamma} a F_{\mu\nu}\tilde{F}^{\mu\nu} - g_{aee} a \bar{\psi}_{e}\gamma_{5}\psi_{e}
\end{equation}
Where \( F^{\mu\nu} \) represents the electromagnetic field strength tensor and its dual is \( \tilde{F}^{\mu\nu} = \epsilon^{\mu\nu\rho\sigma} F_{\rho\sigma} \), $a$ is the axion field, $\psi_{e}$ is the electron Dirac fermion. The parameters \(g_{a\gamma\gamma}\) and \(g_{aee}\) denote the axion-photon and axion-electron couplings, respectively, both expressed in units of \(\mathrm{GeV}^{-1}\)~\cite{aguilar}. 

\begin{figure}[h!]
  \centering
    \includegraphics[width=\linewidth]{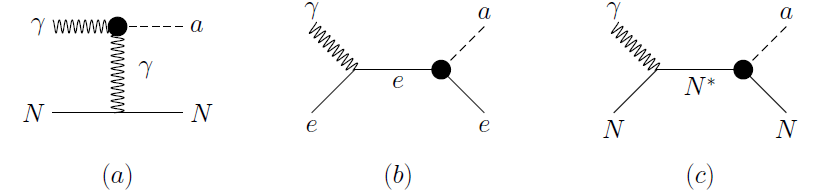}
   \caption{Mechanisms for producing ALPs at the reactor facility: (a) represents the Primakoff process ($\gamma$ + N $\rightarrow$ a + N, where N is the atomic target), (b) represents the Compton-like process ($\gamma$ + $e^{-}$ $\rightarrow$ a + $e^{-}$) and (c) represents nuclear de-excitation ($\gamma$ + N $\longleftrightarrow$ N$^{*}$ $\rightarrow$ a + N) \cite{sierra, sahoo}. Black points are vertices where the axion-like particle is coupled to another particle.}
  \label{fig:alp_production_feynmann}
\end{figure}

A reactor generates high-energy photons through fission processes \cite{roos}. These photons can interact with reactor materials, specifically $^{235}$U in our case, leading to the production of ALPs through the Primakoff process, a Compton-like process, and nuclear de-excitation. Fig.~\ref{fig:alp_production_feynmann} illustrates these three photon-induced mechanisms responsible for ALP production within the reactor.

We will focus on a model where the ALP can be produced through photon or electron coupling and calculate the sensitivities for ALP-photon and ALP-electron couplings. 

Once ALPs are produced, they can be detected through photon production at the detection site via the inverse Primakoff process, \( a + N \rightarrow \gamma + N \), where \( N \) represents the atomic target. Additionally, ALPs can interact with electrons through an inverse Compton-like process, \( a + e^{-} \rightarrow \gamma + e^{-} \), resulting in photon production. ALPs may also decay into two photons or an \( e^{-}/e^{+} \) pair within the detector material, depending on their lifetime. These detection channels are illustrated in Fig.~\ref{fig:fig1}.

In this work, we predominantly adopt the minimal ALP assumption, which allows for ALP-SM couplings governing both the production and detection processes of ALPs \cite{dent}. As shown in Fig.~\ref{fig:alp_production_feynmann}(a), the Primakoff process occurs through a $t$-channel virtual photon exchange, with the production rate depending on the coupling strength $g_{a\gamma\gamma}$ \cite{dent}.
The differential cross-section for the production of ALPs via the Primakoff process in the reactor core is given by \cite{Aloni_2019, PhysRevD.34.1326}:

\begin{equation} \label{eq:prim_prod}
    \frac{d \sigma^{p}_{P}}{d \cos\theta} = \frac{1}{4} g^{2}_{a\gamma\gamma} \alpha Z^{2}F^{2}(t)\frac{|\bar{p_a}|^4 \sin^2 \theta}{t^2} .
\end{equation}

Where $t = (p_1 - k_1)^2 = m_a^2 + E_\gamma(E_a - |\bar{p_a}|\cos\theta)$ is the square of the four-momentum transfer, $\alpha=\frac{e^{2}}{4\pi}$ is the standard electromagnetic fine structure constant, $Z$ is the atomic number of the target nucleus, $F(t)$ is the form factor, $|\bar{p_a}|$ is the magnitude of the ALP's outgoing three-momentum at the angle $\theta$ relative to the incident photon momentum, $E_\gamma$ is the incident photon energy, and $E_a$ is the energy of ALP . It is important to note the production cross-section is enhanced by a coherency factor \( Z^2 \), which increases the sensitivity of our experiment to ALP production and detection through the Primakoff channel compared to other processes. 
The differential cross-section for the inverse Primakoff process follows the same form as in Eq.~(\ref{eq:prim_prod}), with the front factor changing from \( \frac{1}{4} \) to \( \frac{1}{2} \), since the initial spin states involve a spin-0 ALP instead of a spin-1 photon.
The ALP event yield via the inverse Primakoff process is estimated by the convolution of the detection rate with the reactor photon flux as follows \cite{dent}:

\begin{equation} \label{eq:alp_detection_1}
    S(scattering) = \Delta t \frac{N_T}{4\pi l_d^2}
    \int dE_\gamma \sigma_P^d\frac{dN_\gamma}{dE_\gamma}\frac{\sigma_P^p}{\sigma_P^p+\sigma_{SM}}P_{surv}^\gamma .
\end{equation}

Where \( \Delta t \) is the data-taking time, \( N_T \) is the number of target atoms, \( \sigma_P^d \) is the Primakoff scattering cross-section in the detector, \( \frac{dN_\gamma}{dE_\gamma} \) represents the differential photon flux at the detector ($time^{-1}Energy^{-1}$), \( \sigma_P^p \) is the total Primakoff axion production cross-section, \( \sigma_{SM} \) represents the total photon scattering cross-section against core material taken from the Photon Cross Sections
 Database \cite{XCOM} and \( P_{\text{surv}}^\gamma \) denotes the ALP survival probability, accounting for the likelihood that ALPs travel from the reactor core to the detector without decaying:

\begin{equation} \label{eq:alp_surv_prob}
    P_{surv}^\gamma = \exp\bigg(-\frac{l_d}{v_a\tau_a^\gamma}\bigg) .
\end{equation}

Where, $v_a$ is the axion velocity, \( l_d \) is the core-detector distance, and $\tau_a^\gamma = \frac{E_a}{m_a}\frac{1}{\Gamma (a \rightarrow \gamma\gamma)}$ is the axion decay lifetime.

Similarly, for the decay of an ALP into two photons, the signal is estimated as follows \cite{dent}:

\begin{equation} \label{eq:alp_detection_2}
    S(decay) = \Delta t \frac{N_T}{4\pi l_d^2}\int dE_\gamma \sigma_P^d \frac{dN_\gamma}{dE_\gamma}\frac{\sigma_P^p}{\sigma_P^p+\sigma_{SM}}P_{surv}^\gamma P_{decay}^\gamma .
\end{equation}

Where A is the detector transverse area, and $P_{decay}^\gamma$ is the axion decay probability inside the detector volume:

\begin{equation} \label{eq:alp_decay_prob}
    P_{decay}^\gamma = \exp\bigg(-\frac{l_d}{v_a\tau_a^\gamma}\bigg) \bigg[ 1 - \exp\bigg( -\frac{\Delta l}{v_a\tau_a^\gamma}\bigg) \bigg] ,
\end{equation}

and $\Delta l$ is the fiducial detector length. Note that the axion decay width in its rest frame, given by \(\Gamma (a \rightarrow \gamma\gamma) = \frac{g_{a\gamma\gamma}^2 m_a^3}{64\pi}\), scales as \(m_a^3\). Consequently, heavier ALPs are more likely to decay, even for smaller values of the coupling constant, so \(g_{a\gamma\gamma}\) typically decreases with increasing \(m_a\). As the decay lifetime is inversely proportional to the decay width, this leads to a rapid decrease in sensitivity to light axions whose decay length exceeds the experimental flight path. Consequently, heavier axions are more constrained by decay limits \cite{Verma}.
Considering Eqs.~(\ref{eq:prim_prod}) and (\ref{eq:alp_detection_1}), we obtain a relation for the ALP signal rate through scattering (decay processes). An analogous analysis is performed for the axion-electron coupling, where the contribution from atomic ionization via axion absorption is disregarded in the present work to provide preliminary results. However, this contribution will be incorporated in future studies to refine the sensitivity estimates~\cite{Derevianko_2010}. Following the approach in Ref.\cite{dent}, we can calculate axion production via Compton scattering and estimate the corresponding signal rate through inverse Compton scattering. In the following section, we will discuss the data acquisition process and the background count rate, which set a lower limit on the ALP signal rate. This allows us to determine the sensitivity to the coupling $g_{a\gamma\gamma}$ and $g_{aee}$ as a function of \( m_a \).

\section{Experimental setup} \label{Experimental}

This experiment is conducted at the Nuclear Science Center (NSC) at Texas A\&M University. This section describes the specific details regarding the reactor, detector, and shielding methods to limit the backgrounds. We will also describe the data acquisition process and event reconstruction at the end. 

\subsection{Reactor (gamma source)}

\begin{figure}[h!]
\centering
    \includegraphics[width=9cm, height=5cm]{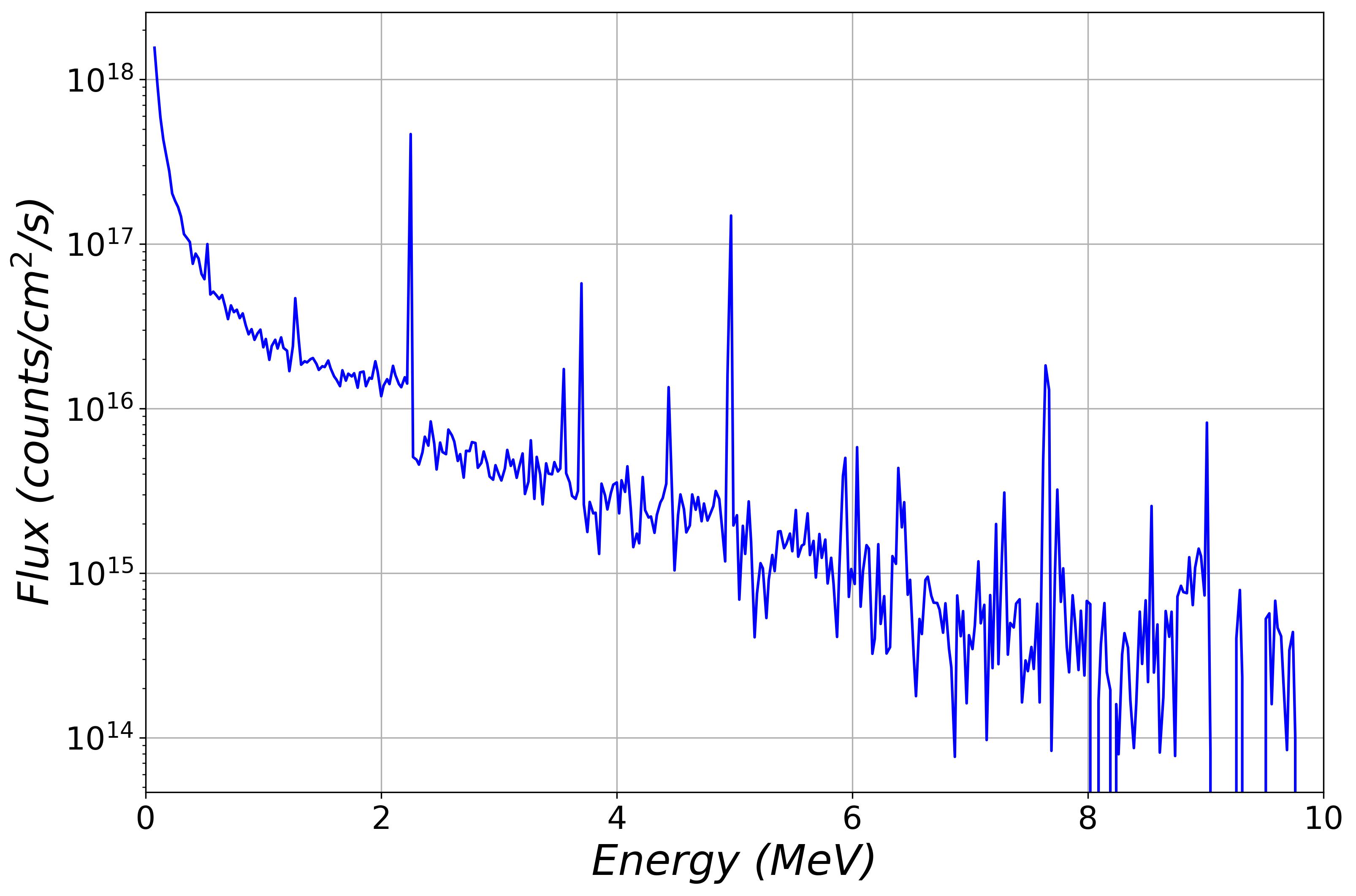}
   \caption{Simulated photon flux at a distance of \( 4\,\mathrm{m} \) from the reactor core, obtained using the MCNP core model. Some of the strongest capture gamma lines in thorium-based reactors are evident at: 2.6 MeV (one of the strongest transitions in \(^{233}\text{Th}\)), 4.0 MeV (the de-excitation of higher-energy nuclear states), and 4.8\textendash 5.0 MeV (a common gamma line from neutron interactions in thorium \cite{AGNOLET201753}).}

  \label{Flux_Reactor_Photon}
\end{figure}

The reactor at the Nuclear Science Center (NSC) is of the TRIGA type (Testing, Research, Isotopes, General Atomics) and can operate at powers up to 1\,MW$_\mathrm{th}$ using low-enriched \({}^{235}\mathrm{U}\) as fuel. This reactor is equipped with a movable core, allowing detectors to be positioned in close proximity to the core, down to a distance of 2\,m. The present experiment was conducted with the reactor operating at an actual power of approximately 800\,kW and a core-to-detector distance of about 4\,m. During operation, the core produces a substantial photon flux of \(\approx 10^{10}\,\mathrm{cm^{-2}\,s^{-1}}\), with an endpoint energy around 10\,MeV. Figure~\ref{Flux_Reactor_Photon} shows the photon flux spectrum originating from the reactor core~\cite{TRIGA}. The gamma-ray spectrum is dominated by neutron interactions, leading to emission lines similar to those observed in other \({}^{235}\mathrm{U}\)-based reactors.

\subsection{Detector Payload}
\begin{figure}[h!]
\centering
    \includegraphics[width=0.40\textwidth]{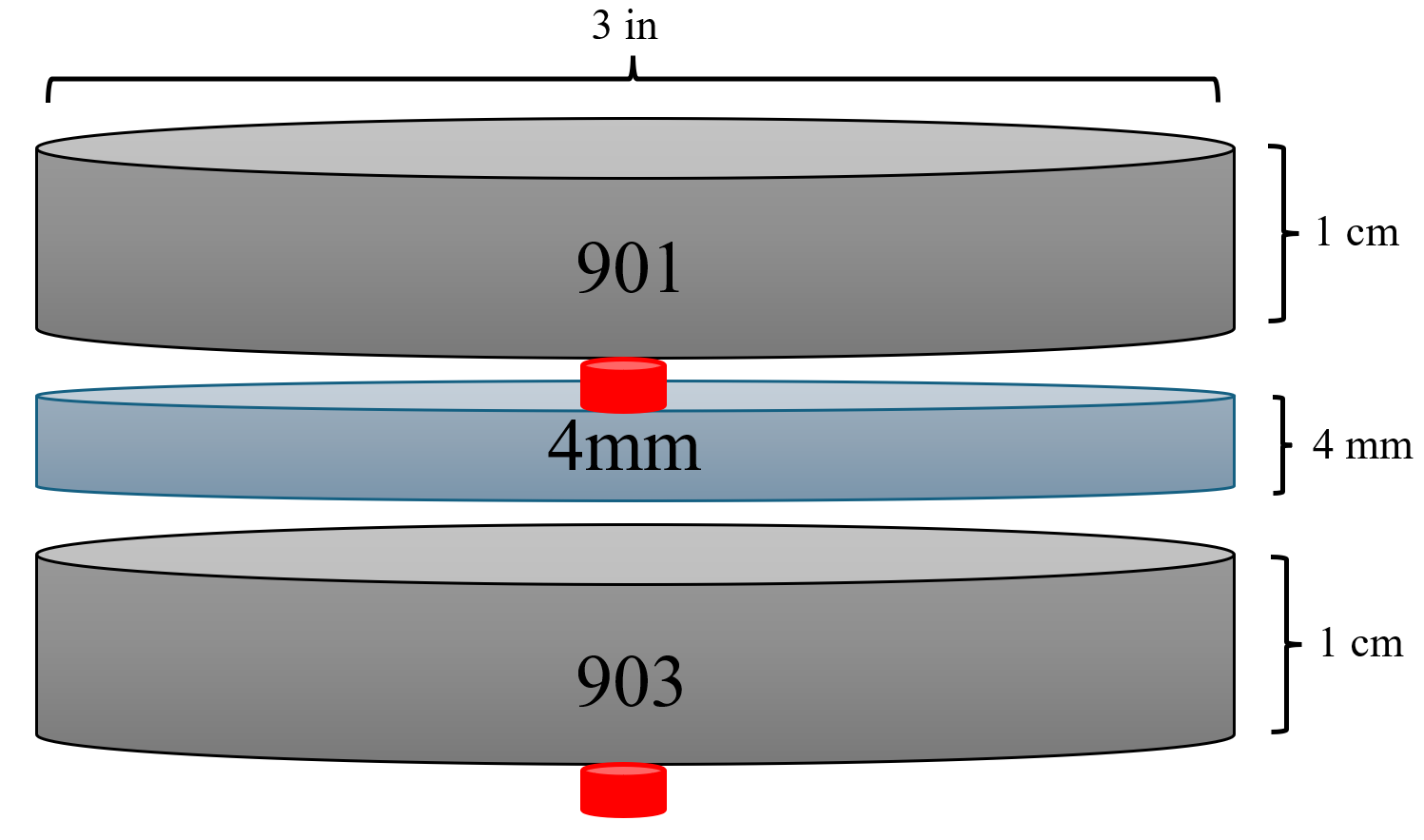}
   \caption{A schematic representation of the three detectors, with the outer detectors labeled 901, 903 (with a thickness 1 cm and mass 181 g) and 4mm (with thickness 4mm and mass 72 g), stacked in a tower configuration with a gap 2 mm between each of them.  For calibration purposes, two \({}^{55}\mathrm{Fe}\) sources are placed between the 901 and 4mm detectors, and below the 903 detector.}
  \label{detector_schematic}
\end{figure}

This experiment employs three cylindrical sapphire detectors, each with a diameter of 3 inches, stacked vertically in a tower configuration within a copper shield. Two of the detectors have a thickness of 1\,cm, while the middle one has a thickness of 4\,mm with a total mass of 435 g. A schematic representation of this tower setup is shown in Fig.~\ref{detector_schematic}. Sapphire (Al$_2$O$_3$), in its single-crystal form, has a density of approximately 3.98\,g/cm$^3$ and is classified as a low-yield scintillator. 

\begin{figure}[h]
\centering
\includegraphics[width=0.35\textwidth]{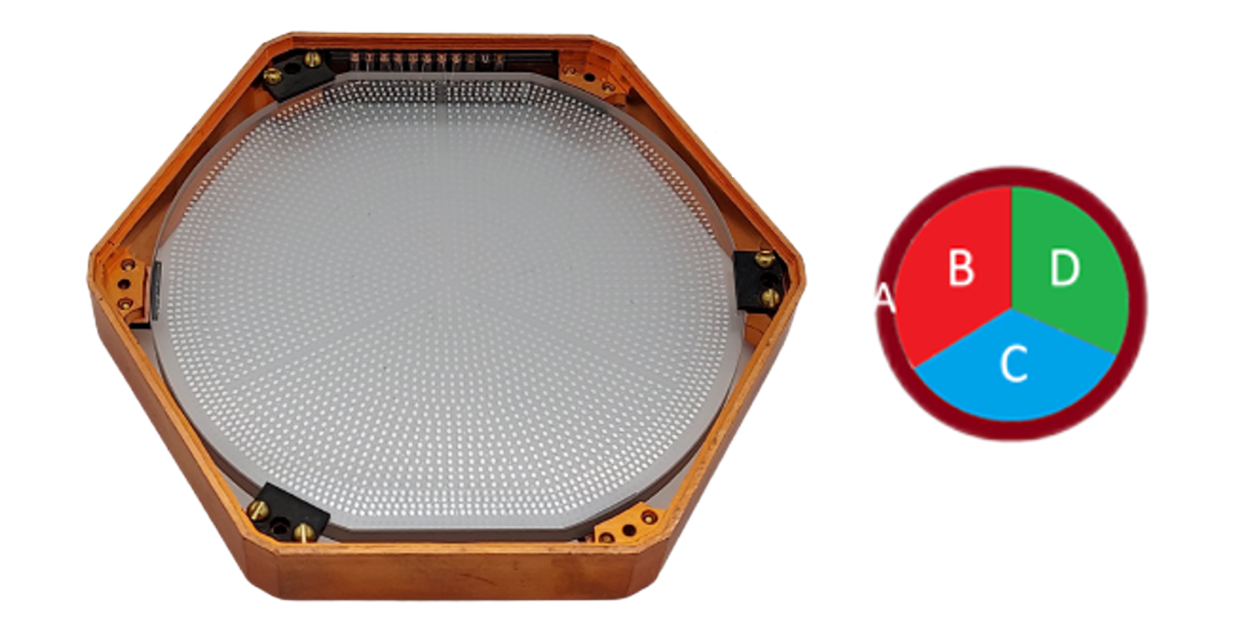}
\caption{ A 4 mm sapphire detector in a copper housing. Reflection
 of the fabricated phonon sensors can be seen due to the transparency of Sapphire crystal. It is equipped with 3 inner
 channels which provide fiducialization for event reconstruction in the bulk of the crystal and an outer channel.} 
\label{fig:detector_channels}
\end{figure}

Each detector is constructed with four readout channels with three inner channels and one outer circular channel as illustrated in Fig.~\ref{fig:detector_channels}. Each channel is instrumented with an array of tungsten Transition Edge Sensors (TES) \cite{10.1063/1.1146105}, which are photolithographically fabricated in parallel on one side of the crystal to measure athermal phonons generated from lattice interactions. The TES are voltage-biased and operate within the superconducting-to-normal transition region. To enable readout through the TES sensors, the detector assembly was installed in a cryogenic system—a BLUEFORS dilution refrigerator—maintained at an operational base temperature of $8\,\mathrm{mK}$ \cite{MALUDZE}. This ultra-low temperature environment is crucial for ensuring the TES sensors operate within their superconducting transition region, thereby achieving the necessary sensitivity to detect athermal phonon signals generated by particle interactions in the detectors.
  

Although the experiment was originally designed for the detection of Coherent Elastic Neutrino-Nucleus Scattering (CE$\nu$NS), it is also well-suited for ALP searches in the low-mass region of the $(g_{a\gamma\gamma}, m_a)$ parameter space. This is enabled by the combination of a substantial photon flux of $10^{10}\,\mathrm{cm^{-2}s^{-1}}$ from the reactor, the close proximity of the detectors to the core, the detectors' low-threshold sensitivity, and the capability to detect ALPs through both scattering and decay channels.  

\subsection{Shielding}
\begin{figure}[h]
\centering
\includegraphics[width=0.35\textwidth]{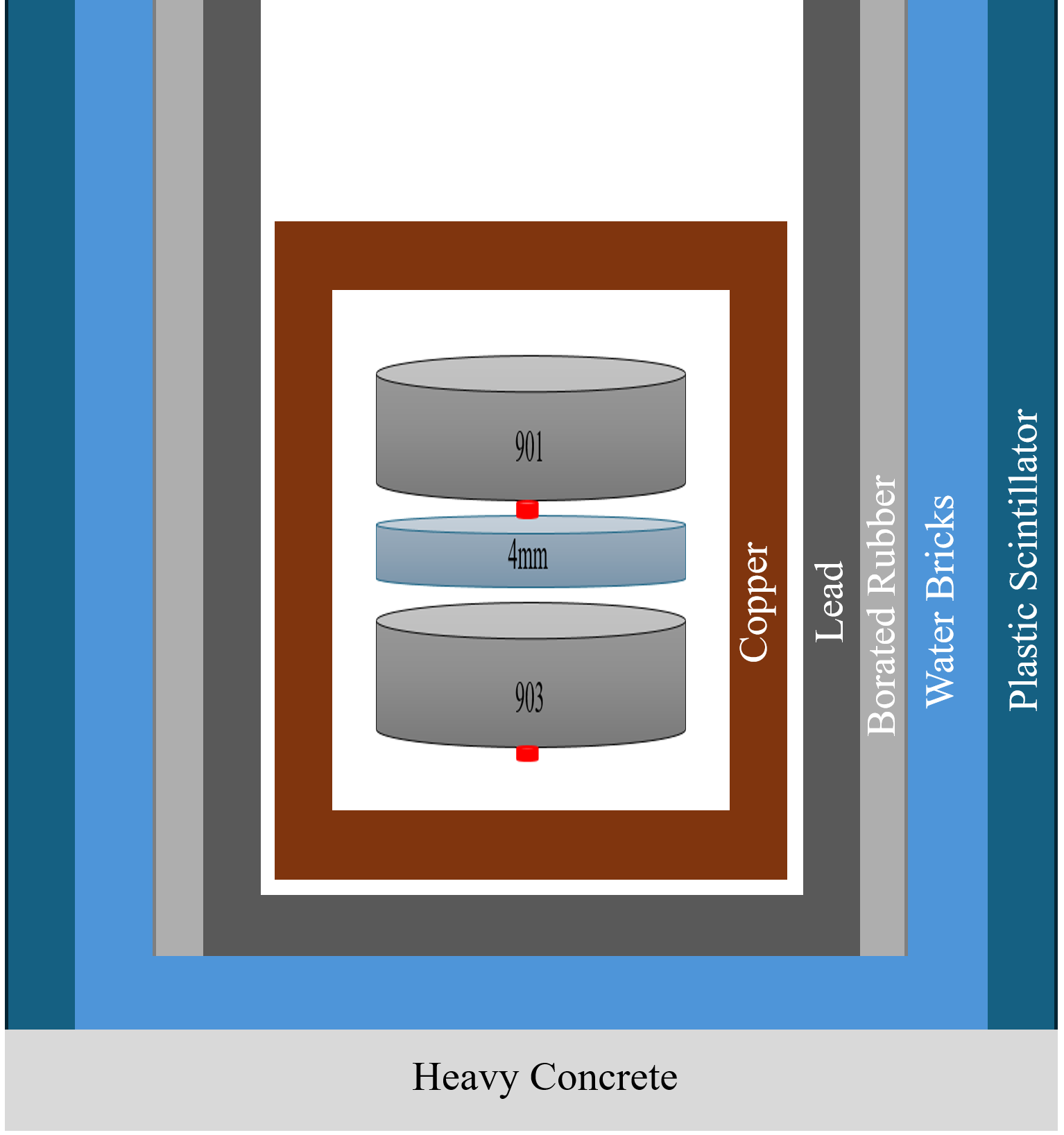}
\caption{ MINER experiment shielding.} 
\label{fig:shielding}
\end{figure}

In this experiment, the detector tower is enclosed within hermetic shielding to suppress the gamma-ray background. To mitigate potential neutron backgrounds, the copper housing is additionally shielded by four layers of lead, a 2\,mm-thick borated rubber layer, and an 8-inch layer of water bricks. Additionally, as a veto and shielding against thermal neutrons, Bicron-412 plastic scintillators with a thickness of 2 inches were placed around the experimental setup.
A schematic of the shielding is shown in Fig. \ref{fig:shielding}.

\subsection{Data acquisition}

For data acquisition, we have used a VME-based CAEN V1740D digitizer which is a 30 MHz analog-to-digital converter with 64-channels, 12-bit resolution, and 62.5 MS/s sampling rate with DPP-QDC (Digital Pulse Processing - Charge to Digital Converter) firmware which can process the input pulses to produce integral (energy) and trigger time stamps for each event in the detector. 
\par The data acquisition process consists of two distinct phases: Reactor-off and Reactor-on. The Reactor-off phase is used to measure background data, providing a baseline for setting a lower limit on the possible ALP signal region and modeling the background spectrum. In the Reactor-on mode, we use its spectrum in setting the exclusion limit to the present work. Given that our setup comprises three detectors stacked in a tower configuration, every detected event must be carefully analyzed to determine its origin. In the following section, we will detail the criteria for defining signal events and the methods used to reject noise.

\section{Analysis and Results} \label{Analysis}
This section describes the definition of single-scatter events, detector calibration and compares resulting spectra with estimated ALP signals to set exclusion curves. 

\subsection{Single-scatter event definition}

Due to the rarity of ALP-induced events, only events detected by a single detector are considered as possible ALP events which are called here as single-scatter events. The focus on these events is motivated by the expectation that multiple-scatter events (i.e. those involving more than one detector) are predominantly associated with background sources such as neutrons, which are capable of interacting with multiple detectors. In contrast, ALP-induced events are expected to produce localized, single scatters within a single detector due to their rare and weakly interacting nature. The selection criteria for such events are as follows:
\begin{enumerate}
    \item Each detector is equipped with four readout channels. Any valid energy deposition in a detector must be simultaneously recorded in all four channels. The total deposited energy (the sum over all four channels) must exceed three times the baseline resolution of the respective detector.
    \item A coincidence time window of 200\,\(\mu\)s is applied. Due to electronic limitations, phonon bouncing, pulse rise times, and the time separation between consecutive signals, this window represents a reasonable choice. If an energy deposition is detected in all four channels of a single detector, and no coinciding signal is observed in the other two detectors within this time window, the event is classified as a candidate single-scatter event.

\end{enumerate}

Any event satisfying the above criteria is classified as a single-scatter event. Isolating single-scatter events enhances the sensitivity of the experiment to ALP signals by reducing background contributions. Since multiple-scatter events have already been excluded through the selection criteria, the combined set of single-scatter events from all detectors can be effectively treated as originating from a single detector with a total mass equal to the sum of the individual detector masses.

\subsection{Calibration}

\begin{figure}[h!]
\centering
    \includegraphics[width=\linewidth]{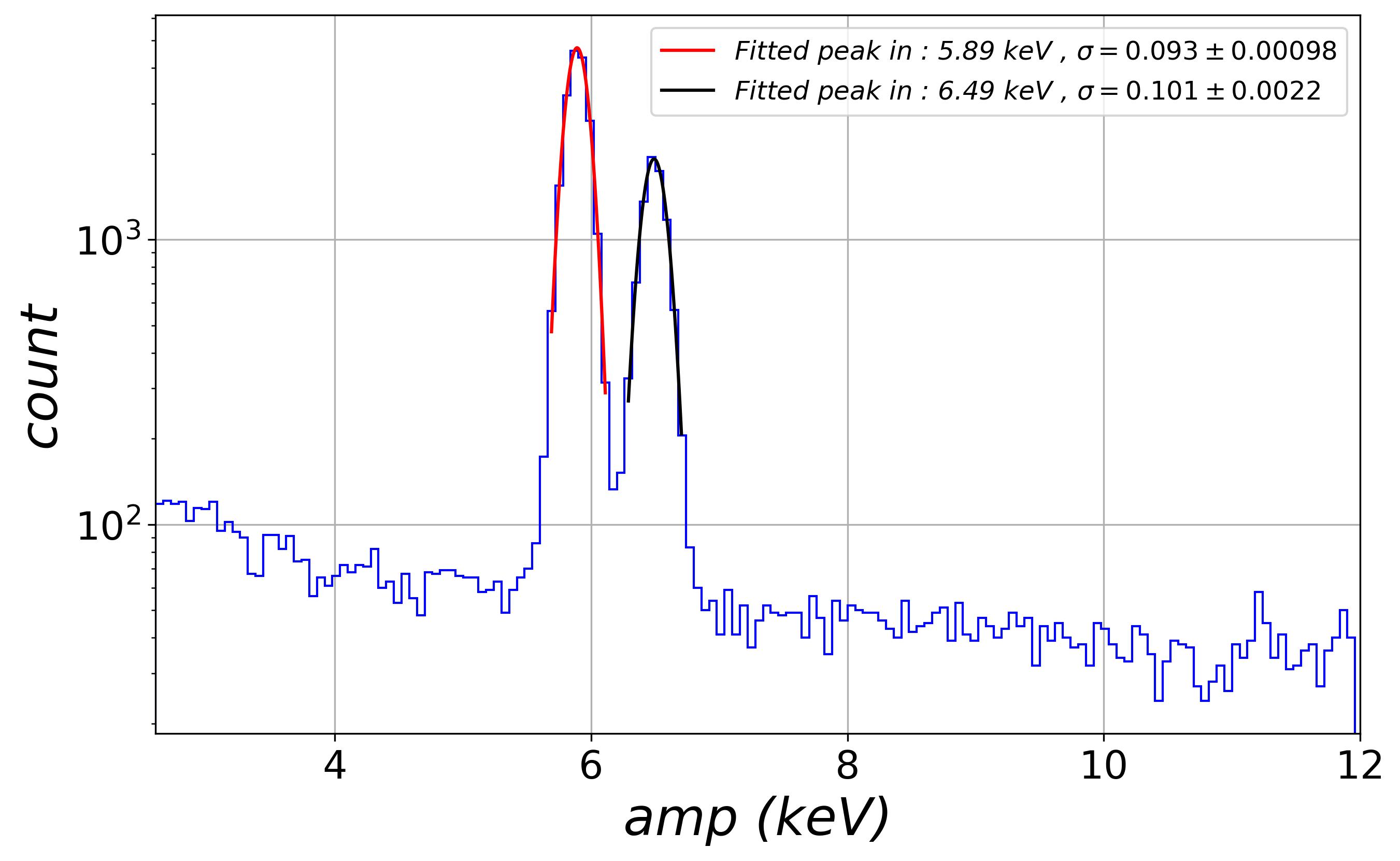}
    \caption{The Optimal Filter (OF) amplitude spectra for 4\,mm detector.}
    \label{fig:calibration}
\end{figure}

The energy spectrum for each detector is obtained by summing the optimal filter amplitudes from its four channels.
  Surface events can be easily ignored. Implementing surface event discrimination based on \(\frac{\text{A}}{\text{ampALL}}\), which represents the ratio of the optimal filter amplitude in channel A to the sum of the amplitudes from all channels for each detector, reject approximately 5\% of events. This demonstrates that the calibration is not affected by surface events and that, to a good approximation, calibrating using all events or using only bulk events yields the same calibration function for each detector.
 Therefore,
Linear calibration for each detector is performed using an ${}^{55}\mathrm{Fe}$ radioactive source, which emits characteristic X-rays of manganese: the dominant \textbf{K\(_\alpha\)} line at approximately 5.89\,keV and the weaker \textbf{K\(_\beta\)} line at approximately 6.49\,keV. To demonstrate the performance of our detectors in resolving the two distinct peaks corresponding to the dominant X-ray emissions from the calibration source, Fig.~\ref{fig:calibration} presents a zoomed-in view of the spectra around the main peaks for 4\,mm detector.
As shown in this figure, this detector exhibits excellent capability in resolving the two distinct peaks from the radioactive source.

The overall baseline energy resolution for each detector after calibration, determined by fitting a Gaussian function to the sum of noise spectrum of all channels, is found to be $45.63 \pm 0.19$\,eV
 for the 4\,mm detector, $140.39 \pm 0.72$\,eV for detector 901, and $126.41 \pm 0.61$\,eV for detector 903. The baseline resolution per channel for the 4\,mm detector is approximately 35 eV, providing a 105 eV low-threshold detector. In the present analysis, the ALP-induced signal was treated as an unsmeared (delta-function-like) contribution within each energy bin. This approximation is justified based on a comparison between the detector energy resolution and the bin width used in the analysis.

The measured detector energy resolution, characterized by \(\sigma\), was found to range between 35\,eV (the baseline resolution per channel) and 101\,eV (the resolution of the second peak in Fig.~\ref{fig:calibration}). Meanwhile, the energy bin width used for signal extraction was 0.4\,keV. The ratio of bin width to energy resolution thus exceeds much (a factor of four) throughout the range, ensuring that the majority of an ALP-induced signal would fall within a single bin. As a result, the impact of energy smearing on the signal distribution is negligible compared to the binning scale.

Neglecting the smearing effect introduces a minimal correction, estimated to be well below the statistical uncertainties of the measurement, and does not affect the sensitivity results at the current level of precision. Therefore, the energy-dependent resolution function, \(\sigma(E) = \sqrt{a^2 + b^2 E}\), where \(a\) accounts for baseline noise and \(b\) describes statistical broadening, was not applied in this work.

\subsection{Experimental results}

The reactor-on data were collected over a total of 59.5 hours across 21 days, between February and September 2022, while the reactor-off data were collected over 163.8 hours across 31 days during the same period. Figure 8 shows the resulting data collect during this run, with events identified as single-scatters shown in the top panel and multiple scatter events shown in the bottom panel for two different phases. The multiple scatter rate was calculated from the total event rate minus the single scatter event rate.

\begin{figure}[h!]
  \centering
    \includegraphics[width=\linewidth]{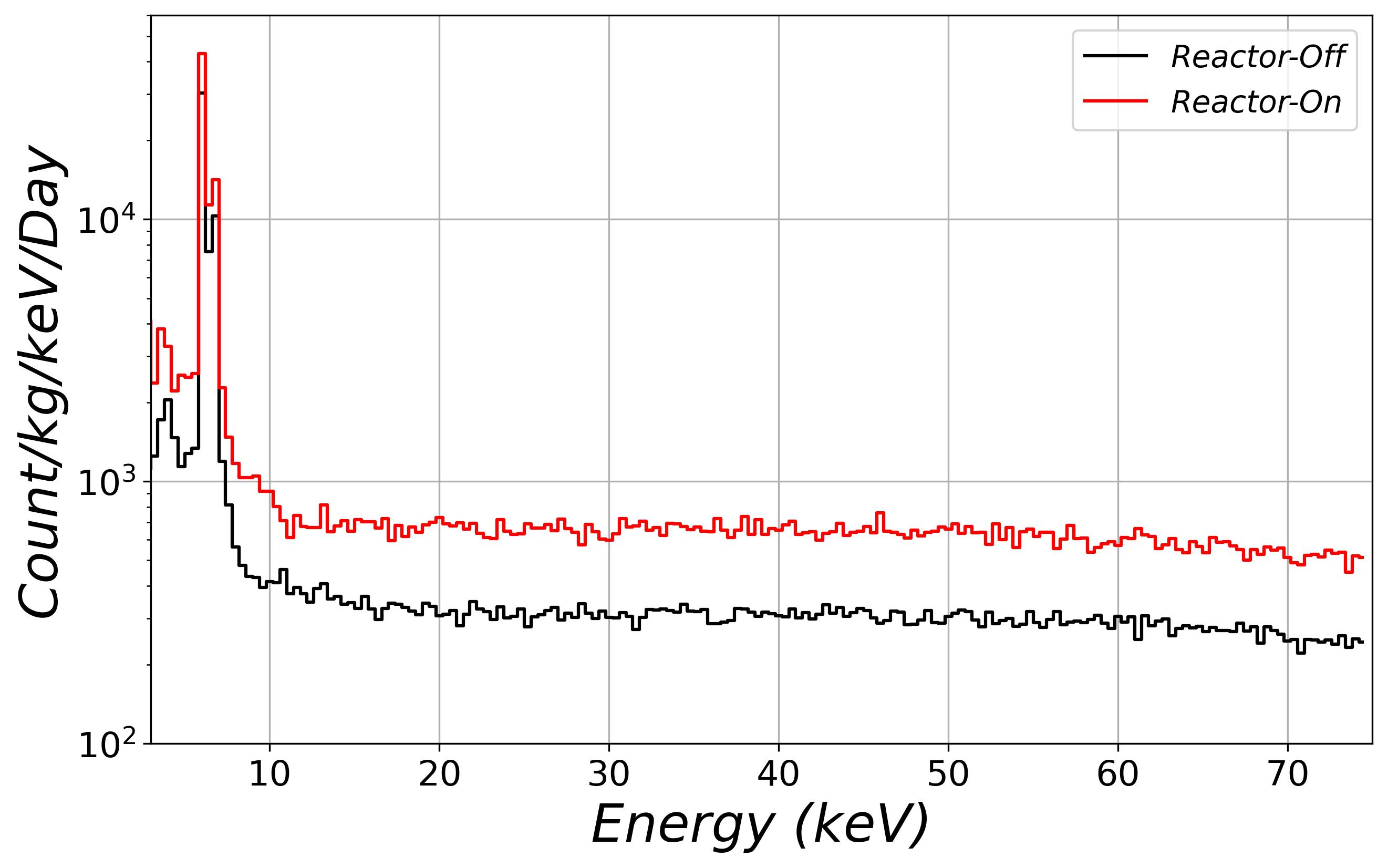} \\
    \includegraphics[width=\linewidth]{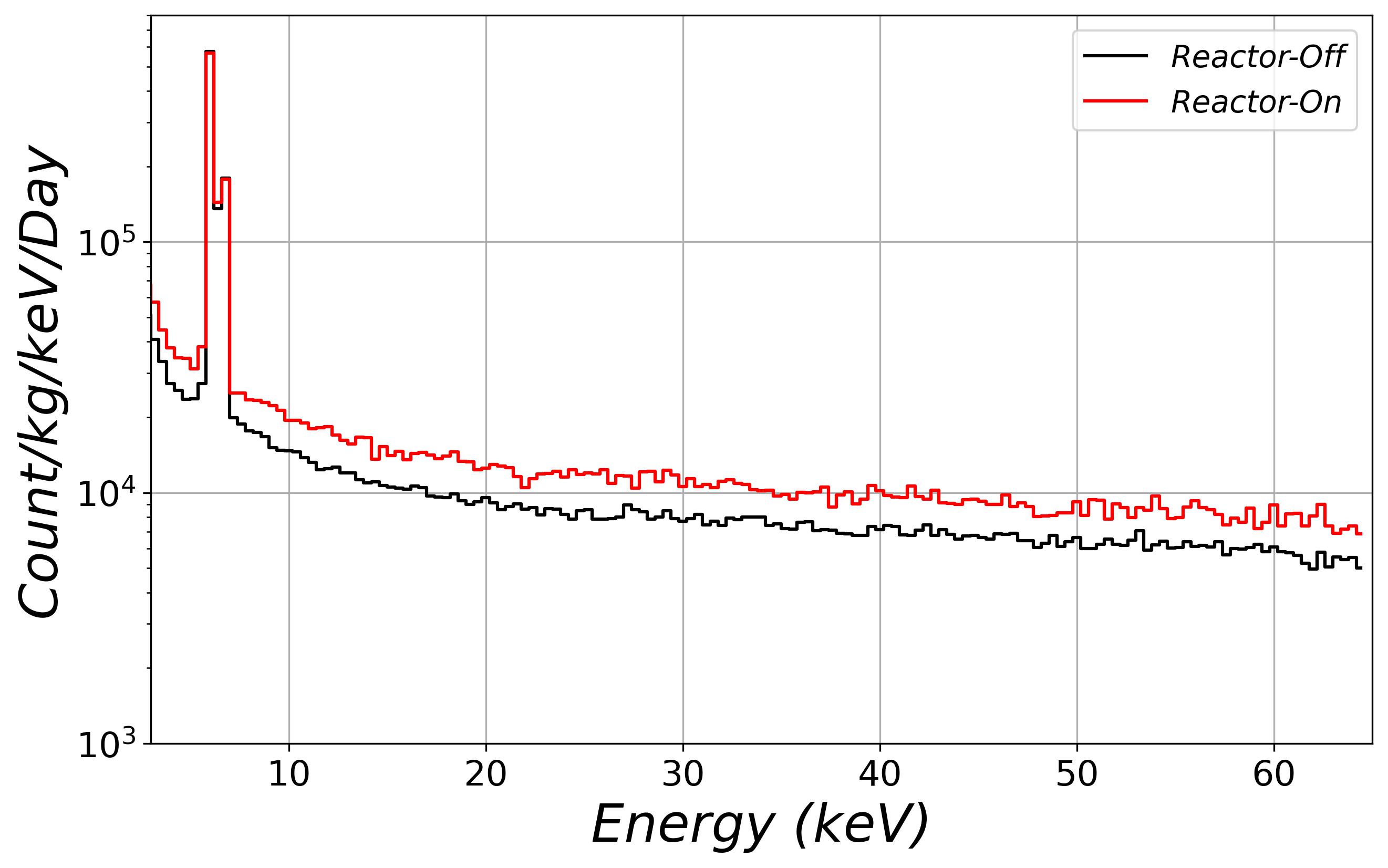}\\
    
   \caption{Measured reactor-on and reactor-off count rates for single-scatter events (top) for combination of 3 detectors and multiple-scatter events (bottom) for 4\,mm detector. These events are identified according to the definition of single-scatter events described in Section~4.1.}
  \label{fig:DRU}
\end{figure}

\subsection{ALP signal rate estimation}

To calculate the ALP production rate we convolve the photon flux with the Primakoff production cross-section and Compton scattering cross-section. We adopted the reactor photon flux from the MINER experiment, as described in~\cite{AGNOLET201753} and shown in Fig.~\ref{Flux_Reactor_Photon}. For projections involving the HFIR reactor, we assume that the integrated photon flux scales linearly with the thermal power of the reactor core. For simplicity, the reactor core material is approximated as pure thorium (\(Z = 90\)), representing an average effective atomic number for the core composition. 

\begin{figure}[h!]
  \centering
    \includegraphics[width=\linewidth]{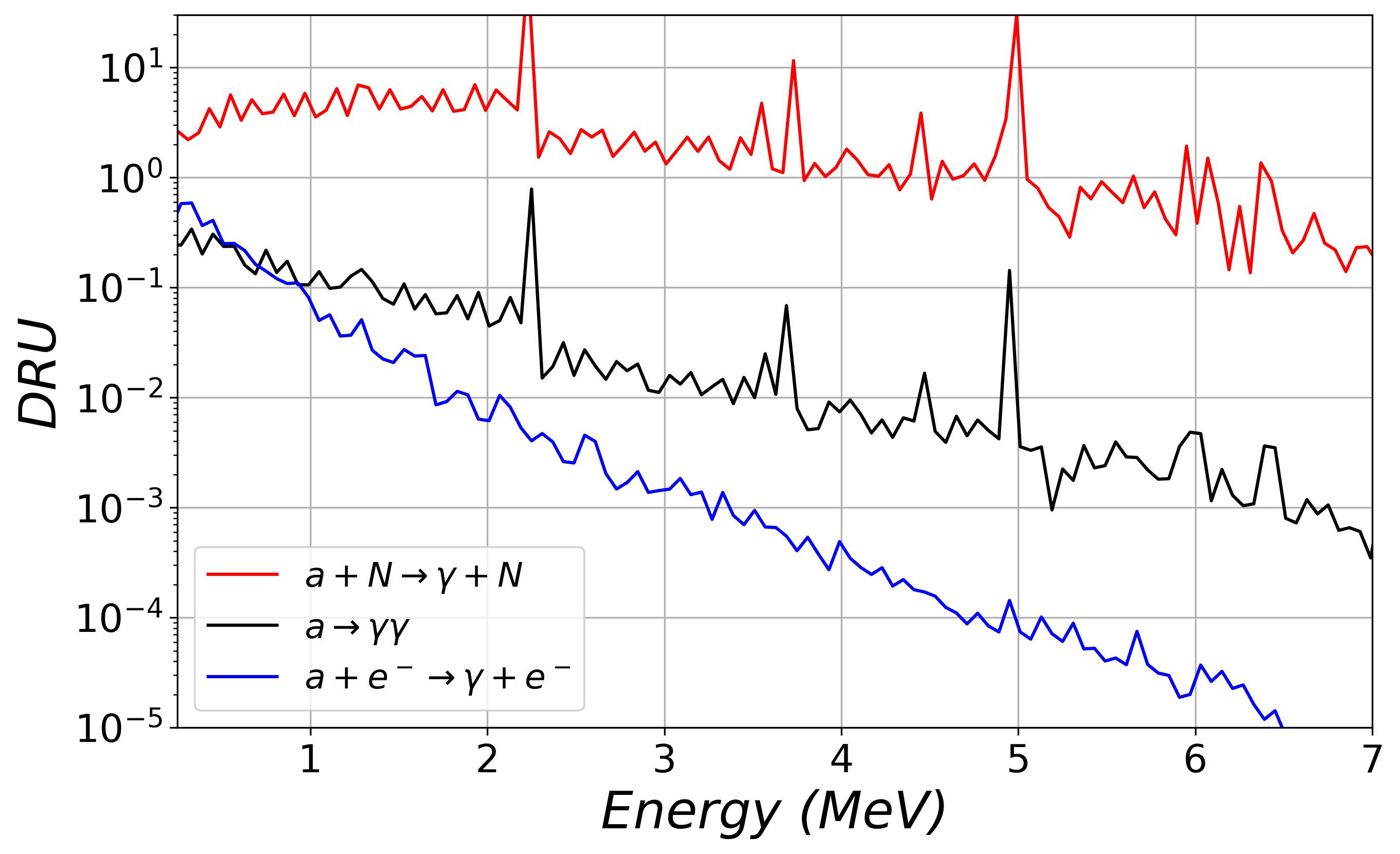}
   \caption{Representative example of photon spectrum resulting from inverse Primakoff scattering, decaying ALPs, and inverse Compton scattering in the detector for a given $(g_{a\gamma\gamma}=10^{-3}, g_{aee}=10^{-5}$ and $m_a=1$~keV ).}
  \label{fig:simulation}
\end{figure}

The produced ALPs then propagate through the shielding and surrounding materials. During propagation, ALPs may produce photons through decay-in-flight, the inverse Primakoff process, or inverse Compton scattering. The observable photon spectrum is the sum of these processes, which are calculated from Eq.~(\ref{eq:alp_detection_1}) and Eq. (\ref{eq:alp_detection_2}) (the equation for inverse Compton scattering is given in ~\cite{dent}). As an illustrative example, by integrating over the photon energy spectrum emitted from the reactor and substituting the relevant experimental parameters — including the detector area (\(A\)), the distance between the detector and the reactor core (\(l_{d}\)), and material-specific properties of the sapphire detector, such as the number of target atoms (\(N_T\)) — Fig.~\ref{fig:simulation} shows the resulting photon spectrum for an assumed ALP mass of \(m_a = 1\,\mathrm{keV}\) and coupling constants \(g_{a\gamma\gamma} = 10^{-3}\)~\(\mathrm{GeV}^{-1}\) and \(g_{aee} = 10^{-5}\)~\(\mathrm{GeV}^{-1}\). These simulated results are in good agreement with the findings reported in Ref.~\cite{dent}, thereby validating the exclusion limits derived from our simulations.
.

The single-scatter background events measured during the reactor-off phase serve as a lower limit on the potential ALP-induced event rate. This limit is subsequently used to constrain the ALP signal rate, as expressed in Eqs.~(\ref{eq:alp_detection_1}) and (\ref{eq:alp_detection_2}). Any genuine ALP signal must exceed the measured count rate of these single-scatter background events. In reactor-off mode, since there is no possibility of ALP production, any observed events are purely background.

\subsection{Exclusion limits and projections}

We calculate the projected upper limits on ALP couplings via a single energy bin analysis using $\kappa = \frac{N_s^{2}}{N_b}$ as a test statistic, where \(N_s\) and \(N_b\) are the integrated signal and background events, respectively. We calculate the upper limit on \(N_s\) with $\kappa = 4.61$ (corresponding to a 90\% Confidence Level) and assuming a background given by the single scatter event rate in reactor-off mode (shown in Fig. \ref{fig:DRU}).

\begin{figure}[h!]
   \includegraphics[width=\linewidth]{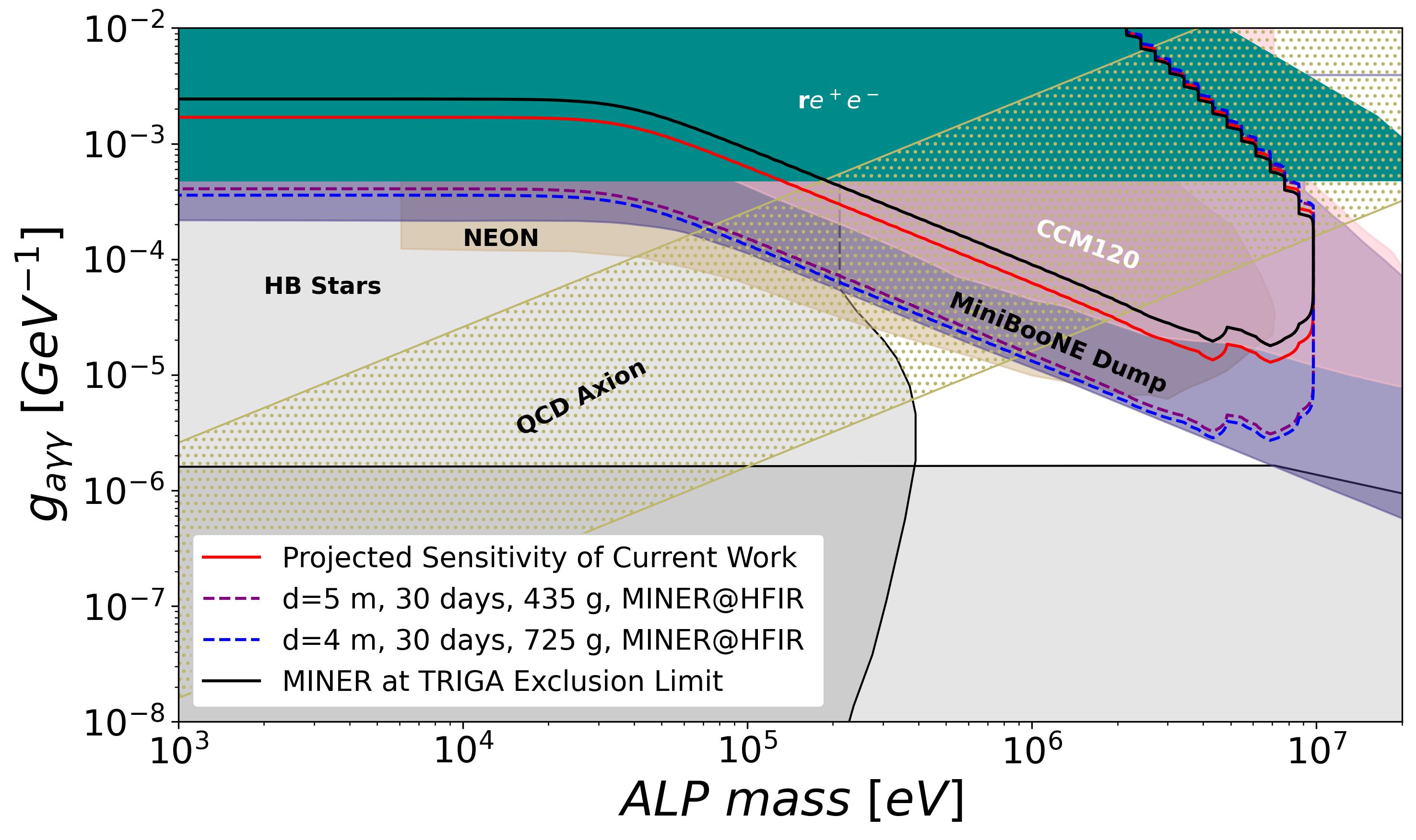}\\
   \includegraphics[width=\linewidth]{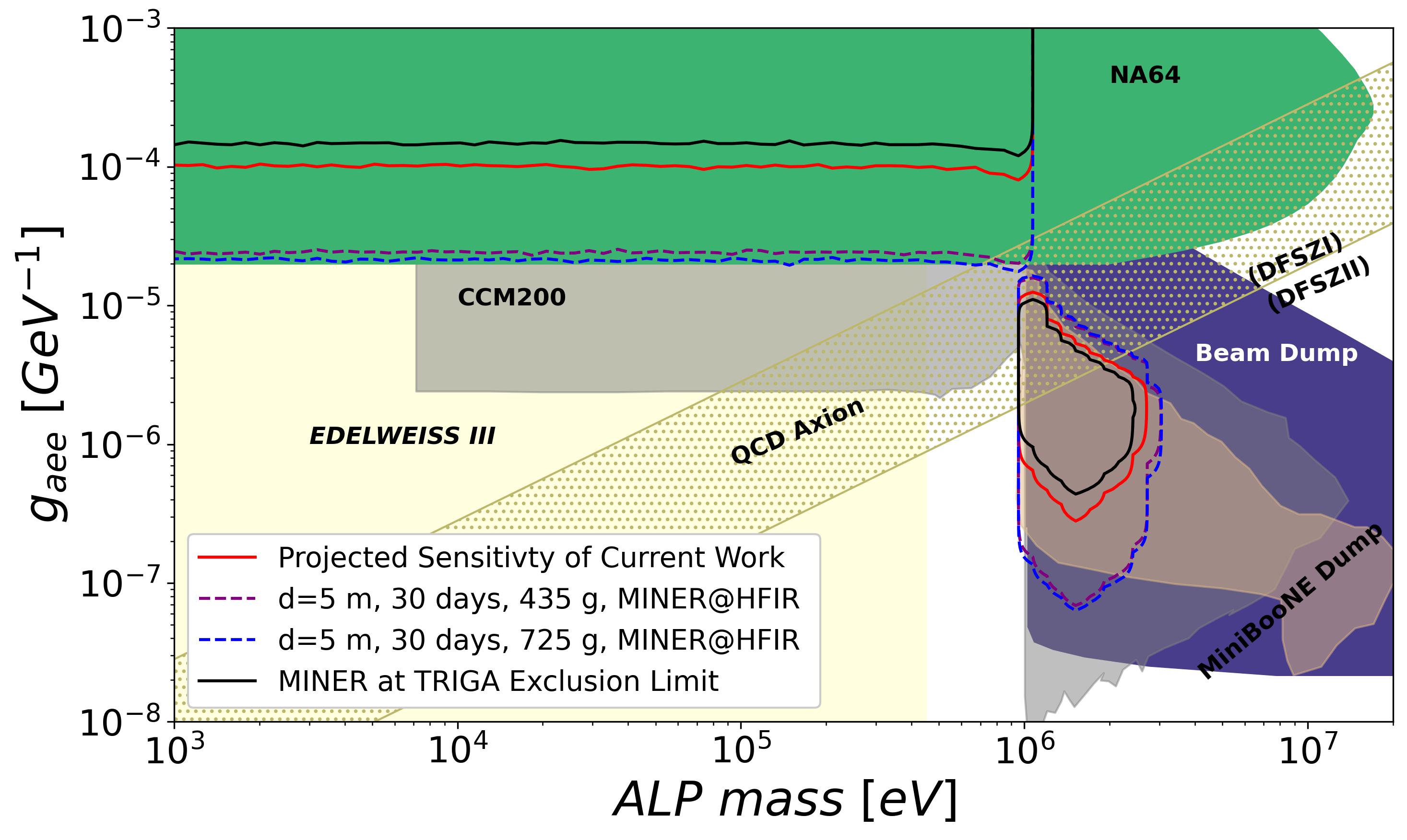}\\
   \caption{Exclusion limits and projected sensitivity to ALP-photon couplings (top) and ALP-electron couplings (bottom) under various experimental conditions.  The black curve shows the exclusion limit obtained from the current experiment, while the red curve corresponds to the achievable experimental sensitivity of the current setup with a lower background. The purple curve represents the projected sensitivity using the combination of all three detectors at the HFIR reactor; and the blue curve depicts the sensitivity achievable with five detectors (totaling 725 g) at the HFIR reactor.}
  \label{fig:sensitivity}
\end{figure}

Fig. \ref{fig:sensitivity} shows $g_{a\gamma\gamma}$ and $g_{aee}$ as a function of the axion mass. In principle, for $m_{a} \lesssim 30\ \text{keV}$, the sensitivity of our setup is independent of the axion mass, resulting in a flat line for the axion-photon coupling. At higher masses, where the decay process becomes dominant, a strong dependence between the coupling and the mass emerges, leading to a downward-sloping sensitivity. For axion-electron coupling, by $\text{1 MeV}$, we expect a flat line for the sensitivity, but due to the pair production after $\text{1 MeV}$ which opens a new decay channel, there is a sharp feature at this region. For higher masses, axion decay is too fast to be detected, which reduces sensitivity.\\

Using reactor-on data and calculated $N_s$, we establish an exclusion limit that our experiment are sensitive to this ALP mass regions. Fig.~\ref{fig:sensitivity} shows the resulting MINER at TRIGA exclusion limit as well as projected exclusion limits at the 85MW High Flux Isotope Reactor (HFIR).

The most remarkable aspect observed in Fig.~\ref{fig:sensitivity}, and a key advantage of our experiment, is the projected sensitivity achievable by scaling from the 1 MWth reactor to the 85 MWth High Flux Isotope Reactor (HFIR), positioned just 5 meters from the detector. As illustrated in Fig. \ref{fig:sensitivity}.(a), this upgrade would allow us to extend our reach into a previously unexplored region of the ALP parameter space (blue and purple curves) in mass region up to 20~keV for $g_{a\gamma\gamma}$, significantly enhancing the potential for ALP discovery. 
For \(g_{aee}\), Fig.~\ref{fig:sensitivity}(b) highlights a remarkable result: the extension of the current detector setup to five detectors — consisting of two detectors with a thickness of 4\,mm and three detectors with a thickness of 1\,cm, for a total mass of 725\,g — would allow access to previously unexplored parameter space, as indicated by the blue curve. Although this configuration was considered in the current experiment, due to technical issues we operated only with three detectors, whose analysis is presented in this work.

\section{Conclusion} \label{Conclusion}
The MINER experiment, employing ultra-low-mass detectors with a low detection threshold and leveraging the high photon flux from reactors, has demonstrated sensitivity to ALP's comparable to much larger experiments. This setup presents an opportunity to probe previously unexplored regions in the ALP parameter space. Looking ahead, the current detector tower can be expanded to accommodate up to five detectors, increasing the total mass to approximately 725\,g. This enhanced setup, in combination with the HFIR reactor at Oak Ridge National Laboratory as the primary photon source, has the potential to significantly improve the projected sensitivity of the experiment. Notably, this upgrade extends our reach deeper into the QCD axion parameter space, a key target for ALP search collaborations.

Other astrophysical constraints are derived from SN1987a and horizontal
branch (HB) stars \cite{BROCKWAY1996439, PhysRevD.33.897, PhysRevD.36.2211, PhysRevLett.113.191302, CARENZA2020135709, Lucente_2020}.
Lab-based limits are provided by beam-dump and collider experiments for high-mass ALPs and low-mass ALP limit is provided by CAST and SUMICO \cite{AristizabalSierra:2020rom}. As seen in Fig.~\ref{fig:sensitivity}, we would reach the experimentally unexplored astrophysical region just by using the high flux photons created byby thehe HFIR reactor and taking data for 30 days, which can extend our current results into this unexplored region. Since this work is limited to the mass range of 3~keV to 10~MeV, we plan to extend our search to the sub-3~keV region in future investigations.

\section{Acknowledgements}
We acknowledge the funding support provided by the Mitchell Institute, the U.S. Department of Energy Detector R\&D Frontier under award numbers DE-SC0017859 and DE-SC0018981, as well as the Department of Atomic Energy (DAE) and the Department of Science and Technology (DST) of India.

\begingroup
\raggedright
\bibliographystyle{elsarticle-num}
\bibliography{ref}
\endgroup

\end{document}